\documentclass[
  ,draft            
  ]
  {aipproc}

\layoutstyle{6x9}

\include{babarsym}
\newcommand{\sqrts} {\ensuremath{\sqrt{s}}\xspace}
\newcommand{\sqrtsp}{\ensuremath{\sqrt{s^\prime}}\xspace}
\def\eeqq     {\ensuremath {e^+e^-\! \to q\overline{q}}\xspace}
\def\eecc     {\ensuremath {e^+e^-\! \to c\overline{c}}\xspace}
\def\Thp      {\ensuremath {\Theta(1540)^+}\xspace} 
\def\Ximm     {\ensuremath {\Xi(1860)^{--}}\xspace} 
 
\def\Thc      {\ensuremath {\Theta_c(3100)^0}\xspace} 
\def\Lc       {\ensuremath {\Lambda_c^+}\xspace} 
\def\dstm     {\ensuremath {D^{*-}}\xspace} 
\def\pdstm    {\ensuremath {p D^{*-}}\xspace}

\begin{document}

\begin{flushright}
\babar-PROC-05/061 \\
SLAC-PUB-11538 \\
\end{flushright}
\vspace*{-1cm}

\title{Hadron Physics at BaBar}

                

\author{David Muller\\ Representing the BaBar Collaboration}
{
  address={Stanford Linear Accelerator Center, Stanford,  CA 94309, USA}
}

\begin{abstract}
The \babar\ experiment at SLAC is designed to measure CP violation in
the $B$ meson system, however the very high statistics combined with
the different $e^+$ and $e^-$ beam energies, the detector design and
the open trigger allow a wide variety of spectroscopic measurements.
We are beginning to tap this potential via several production
mechanisms.
Here we present recent results from initial state radiation, hadronic jets,
few body $B$ and $D$ hadron decays, and interactions in the detector
material.
We also summarize measurements relevant to $D_s$ meson spectroscopy, 
pentaquarks and charmonium spectroscopy from multiple production 
mechanisms.
\end{abstract}

\maketitle

\section{Introduction}

The \babar\ program at SLAC is designed to measure CP violation in the 
$B$ meson
system by producing coherent $B\overline{B}$ pairs and studying the
time dependence of their decays into CP eigenstates.
A 9 \gev electron beam and a 3.1 \gev positron beam collide at a center
of mass energy of 10.58 \gev, corresponding to the peak of the \Y4S
resonance.
The asymmetric \babar\ detector covers about 85\% of the solid angle in the
$e^+e^-$ center of mass (c.m.) frame, 
and is designed to isolate the low multiplicity $B$ decays that occur
in less than 0.001\% of events through a combination of excellent 
charged particle tracking, charged hadron identification and photon detection.

These features are also beneficial for spectroscopy, and the very high
luminosity required for CP violation studies yields very large
samples of hadrons from hadronic jets and bottom ($B$) and charmed
($D$) hadron decays
and allows us to study $e^+e^-$ annihilations at lower \sqrts
through initial state radiation.
Along with high luminosity come high backgrounds, and we are studying
interactions of both beam halo and final state particles with the
beampipe and detector material.
These allow not only a good understanding of the detector itself,
but also studies of $e^\pm$ and long-lived hadron interactions with
materials such as beryllium, tantalum, silicon, carbon and water.

Here we present an overview of our results in the area of hadron spectroscopy.
Details can be found in these proceedings~\cite{bbrtalks}
and in the references.
The first three sections discuss the three main production
mechanisms we have used so far, and present a number of results.
We follow with sections summarizing studies of $D_s$ meson spectroscopy, 
pentaquarks, and charmonium spectroscopy that involve more than
one of these mechanisms.
We have accumulated over 250 fb$^{-1}$ of data so far and expect
at least four times this.
These results thus constitute the beginning of an extremely rich program of
spectroscopy at \babar, and we look forward to many more results in
the future.

\section{Initial State Radiation}

An $e^+e^-$ pair can couple directly to any resonance with
$J^{(PC)}=1^{--}$ and if a decay mode of that resonance can be
reconstructed then the cross section as a function of \sqrts in
the vicinity of the resonance can be used to determine its mass, total
width, \epem width and width to that mode.
We run 90\% of the time at the peak of the \Y4S resonance 
($\sqrts=$10.58~\gev), 10\% below the \Y4S ($\sqrts=$10.54~\gev)
to study non-$B\overline{B}$ backgrounds, and occasionally at other energies.
Our total hadronic cross section yields improved values of the \Y4S mass,
10579.3$\pm$1.2~\mevcc, total width, 20.7$\pm$3.0~\mev, and \epem
width, 0.321$\pm$0.034~\kev~\cite{y4sparams}.

Annihilations at lower \sqrts can be studied via initial state
radiation (ISR) events, in which one of the incoming leptons emits a
photon and subsequently annihilates with the other.
The cross section for this process is the product of the \epem
annihilation cross section at the reduced c.m.\ energy \sqrtsp 
and a well known radiator function;
the latter falls rapidly with increasing photon energy, but the
former increases at low \sqrtsp and becomes dominated by 
low multiplicity resonant contributions.
By detecting the ISR photon, reconstructing a given final state and
requiring overall 4-momentum conservation, we identify
clean samples of ISR events and measure cross sections from threshold
up to $\sqrtsp =$4.5~\gev.
This includes the charmonium region, and we have
improved measurements of the $J/\psi$, and sometimes $\psi(2S)$, 
branching fraction into each mode studied.

We have previously published~\cite{isr34pi} measurements of the
$\pip\pim\pi^0$, $\pip\pim\pip\pim$, $K^+K^-\pip\pim$ and
$K^+K^-K^+K^-$ final states with better coverage than all previous
experiments and comparable or better precision, using 89 fb$^{-1}$ of data.
These states can be studied both in terms of cross section and
internal structure.
The $\pip\pim\pi^0$ final state is dominated by $\omega$ and $\phi$
resonances, and we have improved the world's knowledge of excited
$\omega$ states.
The $\pip\pim\pip\pim$ final state is dominated by the two-body $a_1(1260)\pi$
intermediate state;
the $K^+K^-\pip\pim$ final state shows no significant two-body states,
but a rich three-body structure including $K^*(890)K\pi$, $\phi\pi\pi$,
$\rho KK$ and $K_2^*(1430)K\pi$;
the $K^+K^-K^+K^-$ final state shows no substructure, in particular no
$\phi$ signal.

\begin{figure}
  \includegraphics[height=.225\textheight]{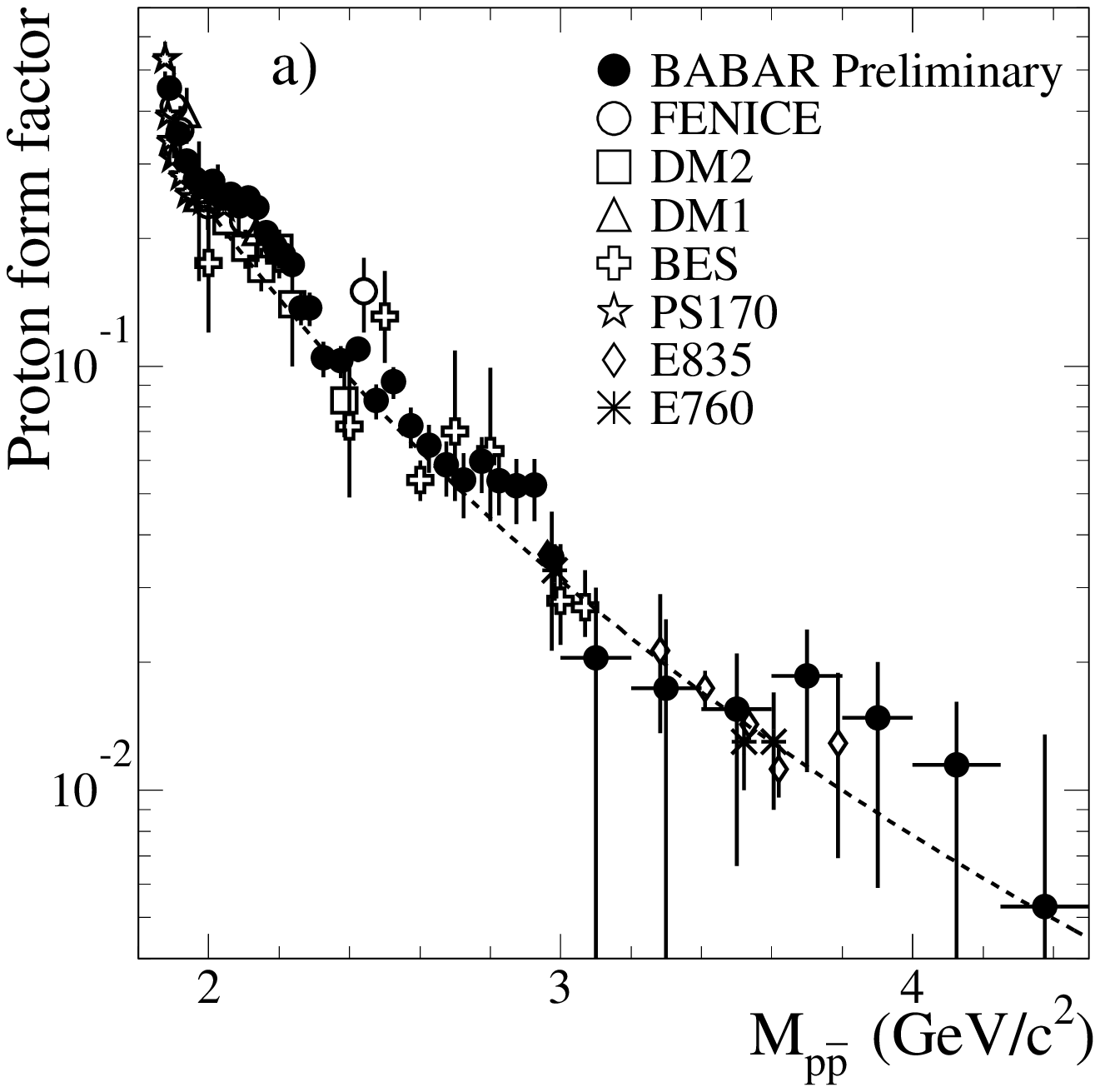}
  \includegraphics[height=.225\textheight]{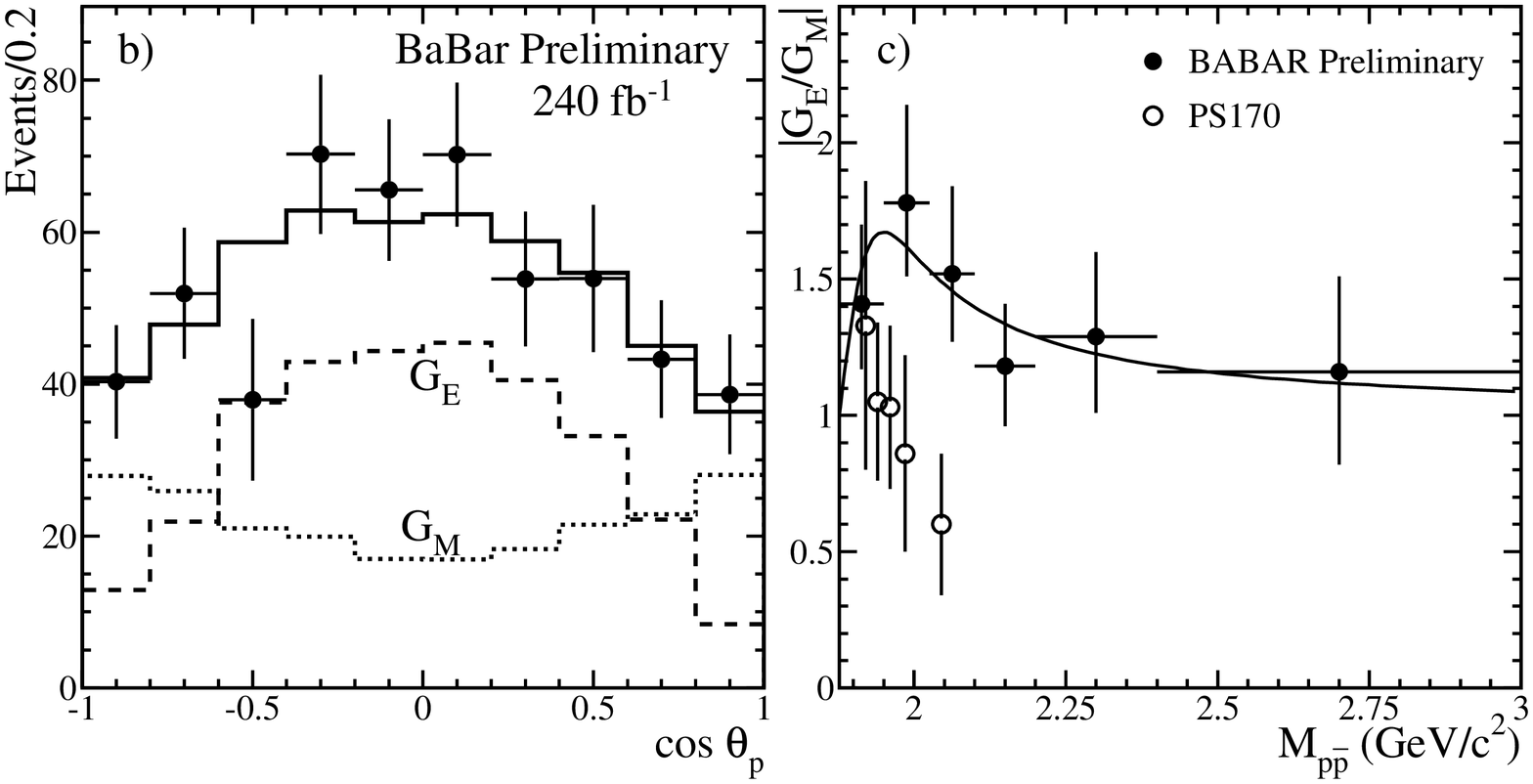}
  \caption{
  \label{ppbarff}
  a) The effective proton form factor as a function of \sqrts derived from our 
  $\epem \to \gamma_{ISR}p\overline{p}$ data, compared with previous results.
  b) The proton angular distribution (points) for events in the range
  1950--2025~\mev; the
  solid histogram represents a fit of the sum of electric (dashed) and
  magnetic (dotted) components.
  c) The ratio of electric and magnetic form factors as a function of \sqrts.
}
\end{figure}

We have a preliminary measurement of the $\epem \to p\overline{p}$ 
cross section;
the corresponding effective form factor is shown in
Fig.~\ref{ppbarff}a, along with previous data from \epem and
$p\overline{p}$ experiments, with which we are consistent.
The \sqrtsp dependence shows the familiar threshold enhancement, as well as
two structures featuring sharp drops at 2.25 and 3~\gev,
which illustrate the power of data from a single experiment over a
wide range with no point-to-point systematic error.
We separate the electric and magnetic form factors using the proton
angular distribution, which is shown in Fig.~\ref{ppbarff}b for
the range $1950<\sqrtsp<2025$~\mev; it deviates significantly from uniformity,
indicating the electric form factor is substantially
larger than the magnetic.
The $G_E$:$G_M$ ratio is shown as a function of \sqrtsp in
Fig.~\ref{ppbarff}c;
the data show a systematic trend toward values above unity at low \sqrtsp,
which is inconsistent with the LEAR data.
The line in the figure is a parametrization used to calculate the
effective form factor in Fig.~\ref{ppbarff}a.

\begin{figure}
  \includegraphics[height=.217\textheight]{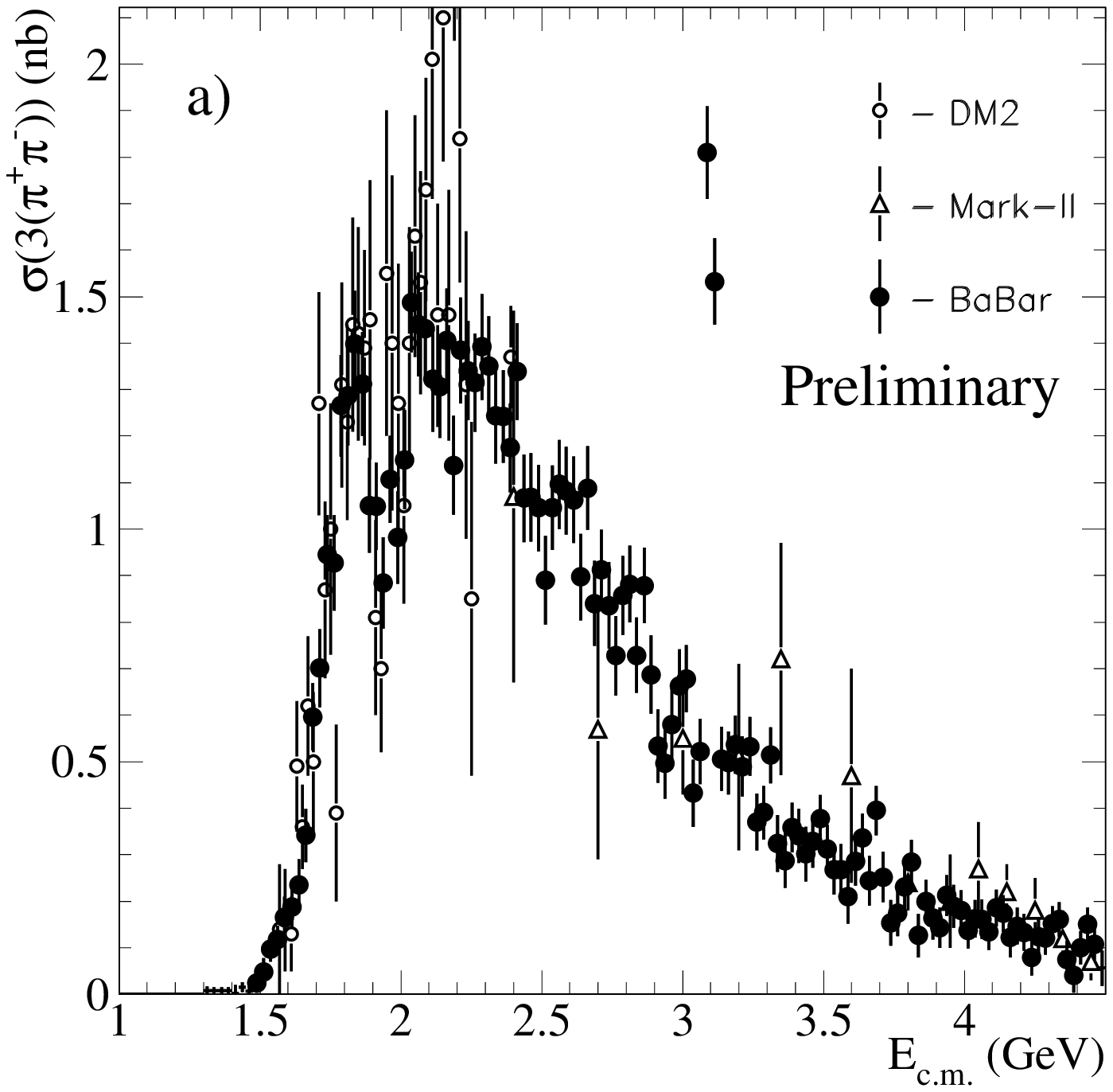}
  \includegraphics[height=.217\textheight]{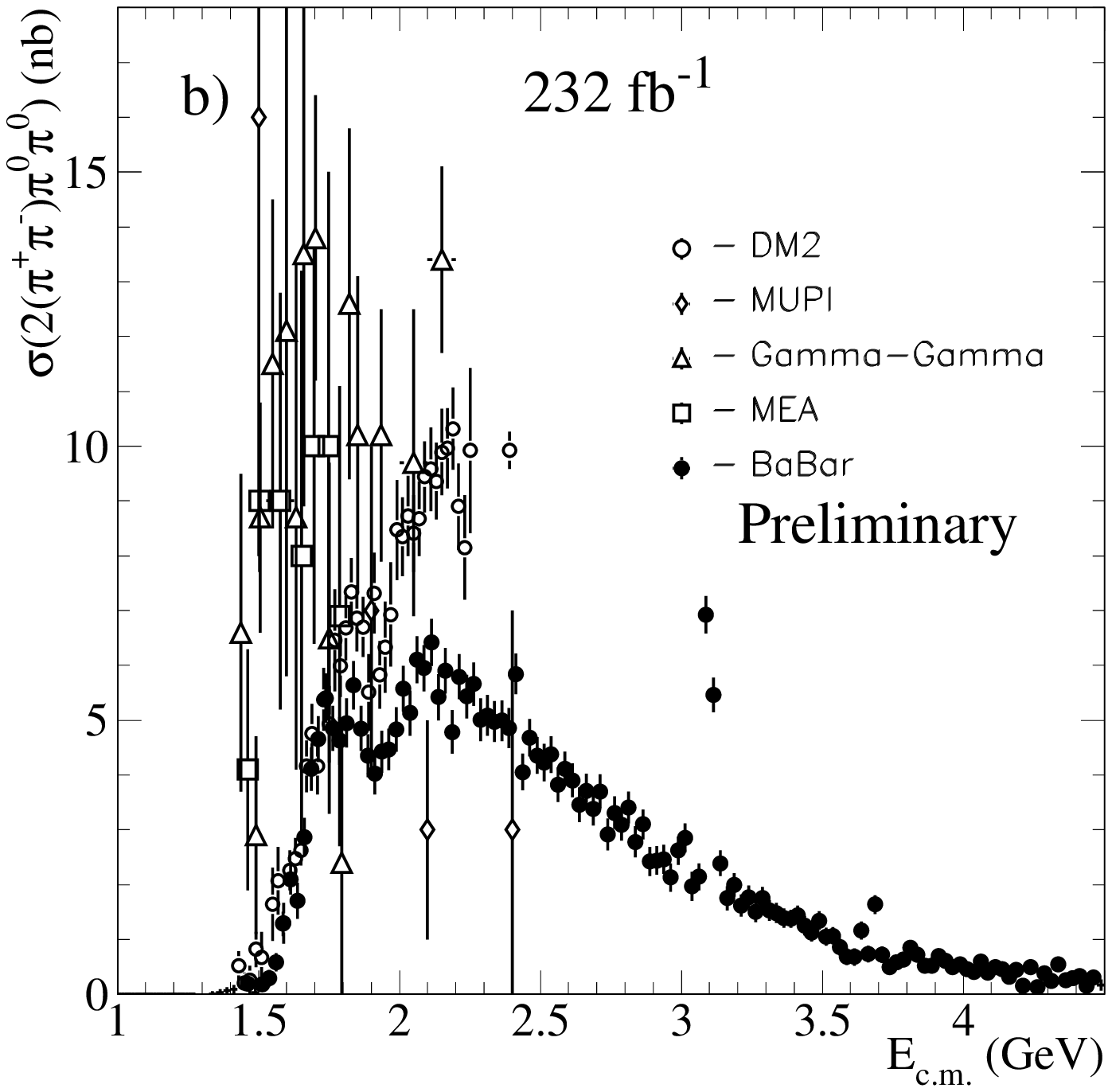}
  \includegraphics[height=.217\textheight]{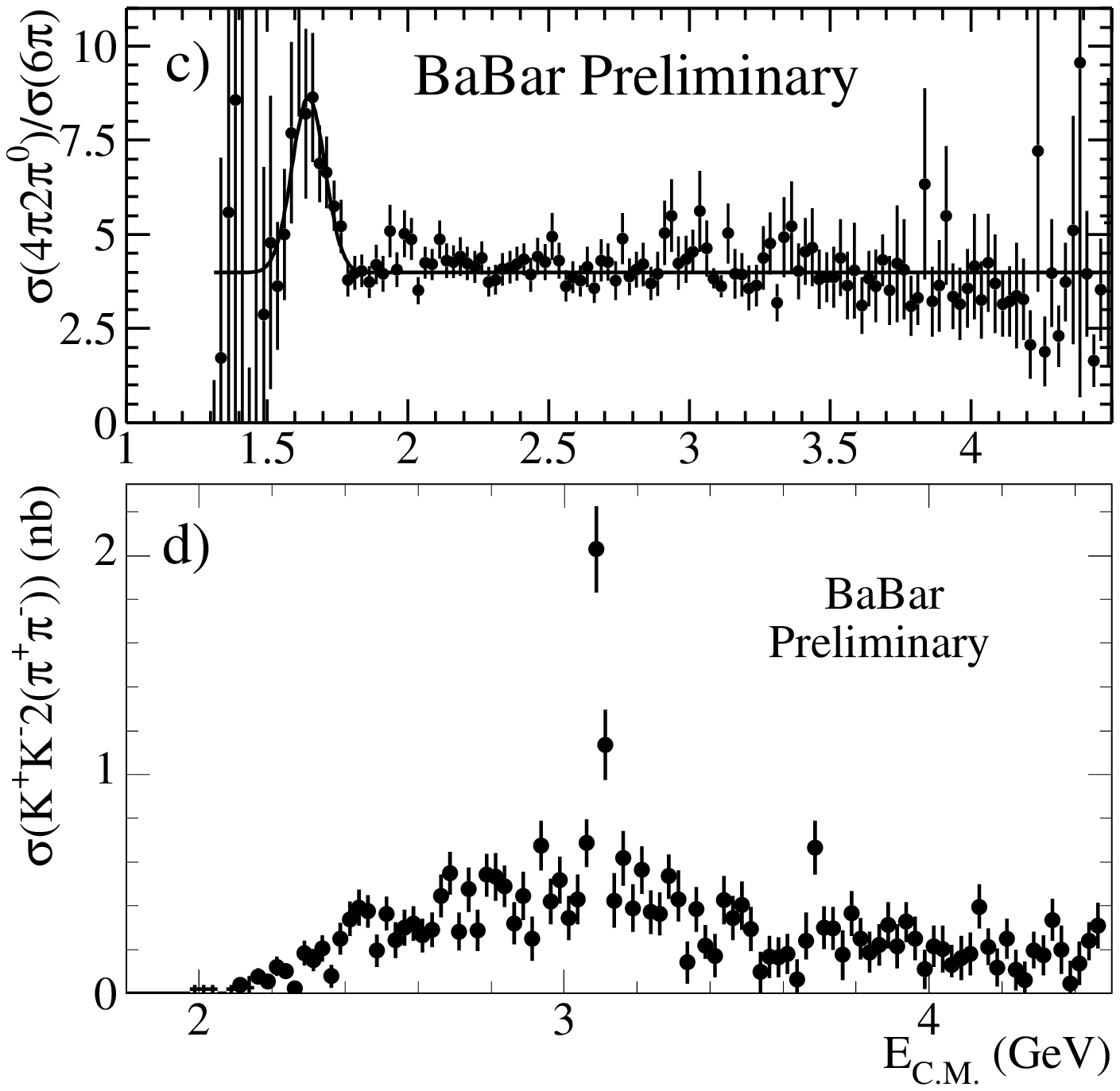}
  \caption{
  \label{6pixs}
  The cross section for $\epem \to$ a) $\pip\pim\pip\pim\pip\pim$ and
  b) $\pip\pim\pip\pim\pi^0\pi^0$ vs.\ \sqrts, compared with previous data.
  c) The ratio of the cross sections in (a) and (b).
  d) The cross section for $\epem \to K^+K^-\pip\pim\pip\pim$.
}
\end{figure}

We also have preliminary studies of the $\pip\pim\pip\pim\pip\pim$, 
$\pip\pim\pip\pim\pi^0\pi^0$ and $K^+K^-\pip\pim\pip\pim$ final states.
The cross sections for the first two are shown in Figs.~\ref{6pixs}a,b;
large improvements over previous measurements are evident.
In the all-charged mode the structure around 1900 \mev seen previously
by DM2 and also in diffractive production by FOCUS, is clear and well
measured.
There is surprisinlgly little substructure in this mode;  
a simulation with one $\rho^0$ and four pions distributed according to
phase space is adequate to describe all the internal distributions.
The $\pip\pim\pip\pim\pi^0\pi^0$ mode shows a rather similar cross
section, but a much more complex internal structure:
we observe signals for $\rho^0$, $\rho^\pm$, $\omega$ and $\eta$, and
a substantial contribution from the two-body $\omega\eta$
intermediate state, which appears to be resonant.
The ratio of the two cross sections is shown in Fig.~\ref{6pixs}c; the
structure near 1650 \mev is due to the $\omega\eta$ contribution, and
the ratio is constant with a value near 4.0 over the remainder of the
range, which is difficult to explain in light of the very
different substructures.
More data and modes containing more $\pi^0$ are under study.

The $K^+K^-\pip\pim\pip\pim$ cross section is measured for the first
time and shown in Fig.~\ref{6pixs}d.
There is interesting substructure, with a weak $\phi$ signal, but
a strong $K^*(890)$.
More statistics and final states such as 
$K^+K^-\pip\pim\pi^0\pi^0$ will be analyzed.

\section{Hadronic Jets}

Our most prevalent events are of the type $\epem \to q\overline{q}$ at 
$\sqrts=$10.58~\gev, in which the quark and antiquark
fragment into a number of hadrons, which produce typically ten stable
charged tracks and photons.
This process is not understood quantitatively, so there is no quantum 
number information from the production process;
however states sufficiently narrow to be seen above background can
yield masses, widths and branching fractions, and
some quantum numbers can be excluded by observed decay modes.

\begin{figure}
  \includegraphics[height=.23\textheight]{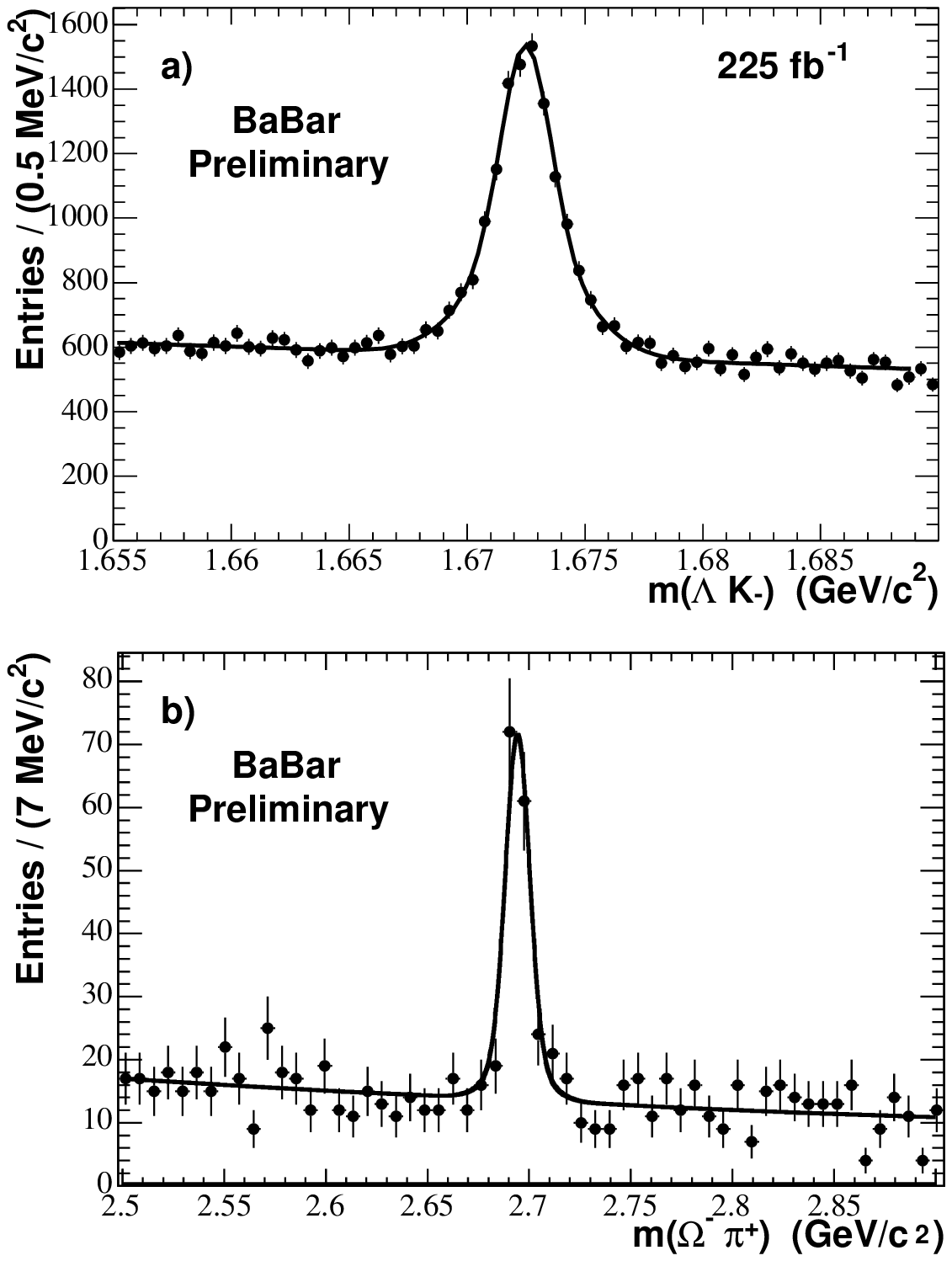}
  \includegraphics[height=.23\textheight]{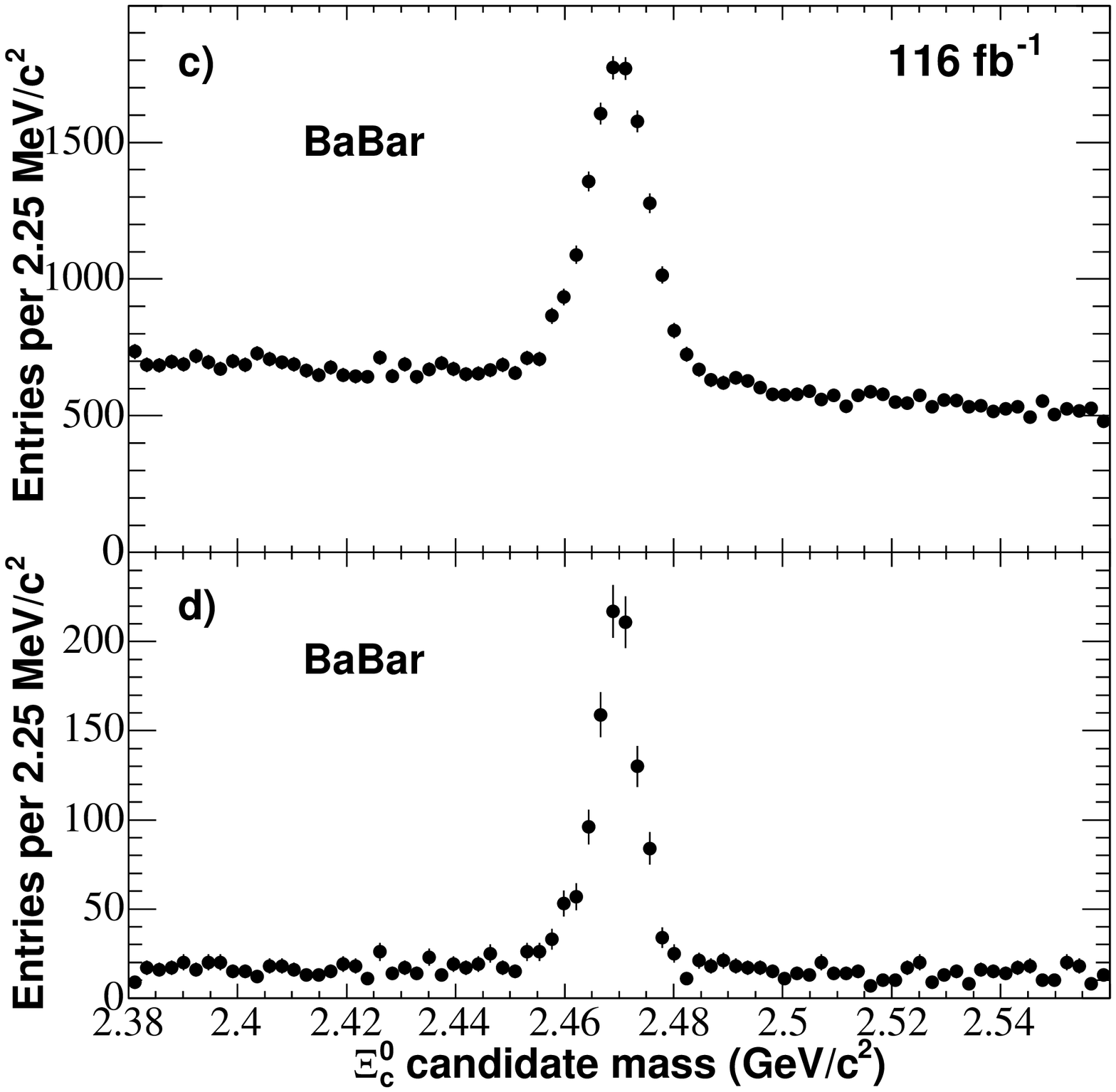}
  \includegraphics[height=.23\textheight]{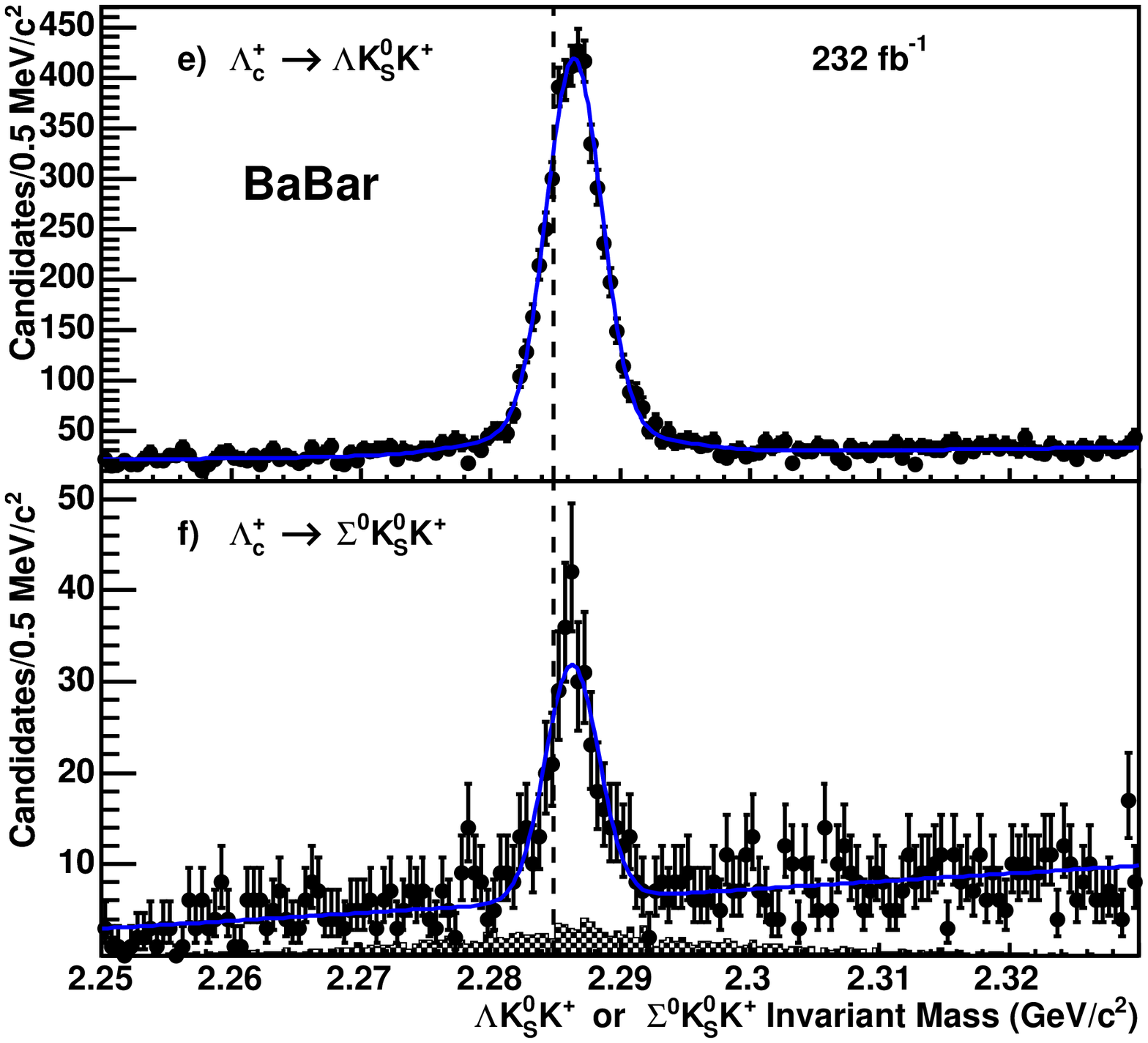}
  \caption{
  \label{cbary}
  Inclusive a) $\Lambda K^-$ and b) $\Omega^-\pip$ invariant mass
  distributions in the vicinity of the $\Omega^-$ and $\Omega^0_c$,
  respectively.
  Inclusive c) $\Xi^-\pip$ and d) $\Omega^-K^+$ invariant mass
  distributions in the vicinity of the $\Xi_c^0$.
  Inclusive e) $\Lambda \KS K^+$ and f) $\Sigma^0\KS K^+$ invariant mass
  distributions in the vicinity of the \Lc.
}
\end{figure}

Our samples of reconstructed hyperons, e.g. $\Lambda^0$, $\Sigma^0$, 
$\Xi^{0,-}$, $\Omega^-$ (Fig.~\ref{cbary}a), $\Xi(1530)$, are much larger
and cleaner than in pre-$B$-factory \epem experiments.
These signals have numerous applications, such as the study of charmed 
baryons from $c\overline{c}$ events and $B$ hadron decays.
For $\Xi^-\pip$ and $\Omega^- K^+$ pairs we show distributions of
invariant mass $M(\Xi^-\pip)$ and $M(\Omega^- K^+)$
in Figs.~\ref{cbary}c and~\ref{cbary}d, respectively;
both show clear signals for the charmed-strange baryon $\Xi_c(2470)^0$
and we measure the branching ratio
${\cal B}(\Xi_c^0\! \to \Omega^- K^+) / 
 {\cal B}(\Xi_c^0\! \to \Xi^-\pip)=0.294\pm0.024$~\cite{xicprl}.
We also observe the doubly strange, charmed baryon $\Omega_c(2700)^0$
in the $M(\Omega^-\pip)$ distribution shown in Fig.~\ref{cbary}b.
Signals for other decay modes are not yet significant, 
but we measure~\cite{omchepex} (preliminary)
\begin{eqnarray*}
 {\cal B}(\Omega_c^0 \to \Xi^-K^-\pip\pip) / 
 {\cal B}(\Omega_c^0 \to \Omega^-\pip) & = & 0.31\pm0.15 \\
 {\cal B}(\Omega_c^0 \to \Omega^-\pip\pim\pip) / 
 {\cal B}(\Omega_c^0 \to \Omega^-\pip) & = & 0.16\pm0.10.
\end{eqnarray*}

We measure the mass of the \Lc baryon using the $\Lambda \KS K^+$ 
and $\Sigma^0\KS K^+$ decay modes.
These have low $Q$ values (and therefore low branching fractions),
which minimize the systematic error due to uncertainties
in the mass scale and resolution of the detector.
The invariant mass distributions for these two modes are shown in
Figs.~\ref{cbary}e and~\ref{cbary}f;  the former mode contains over
4,000 true \Lc baryons and the mass measurement is systematics
limited; 
the latter mode has a smaller systematic error, but is statistics limited.
The masses measured in the two modes are consistent and the combined
result, $M_{\Lambda_c}\! = 2286.46 \pm 0.14$ \mevcc~\cite{lcmassprd},
is four times more
precise than the current PDG value of $2284.9 \pm 0.6$
\mevcc~\cite{pdg} and higher by 2.5 of the PDG standard deviations.

\section{$B$ and $D$ Hadron Decays}

The weakly decaying $B$ and $D$ hadrons are pseudoscalars, with
their (quasi-)two-body decays limited to appropriate quantum numbers 
and helicity states.
These uncommon, low-multiplicity decays can be sensitive to CP-violation, 
and are the goal of our program.
We often use "applied spectroscopy", e.g. using angular distributions
to reduce backgrounds or project out CP-even and CP-odd components.
Recently we improved our measurement~\cite{psiKsprd} of CP
violation in $B^0\!\to J/\psi K^{*0}$ decays, including a full
analysis of the $J/\psi K \pi$ system.
We see phase motion around 890 \mevcc, the direction of which
resolves a sign ambiguity in the cosine of the CP-violating phase, 
and the S-P phase difference 
agrees with corresponding LASS data from $K^-p\!\to\! K^-\pip n$ scattering.

\begin{figure}
  \includegraphics[height=.198\textheight]{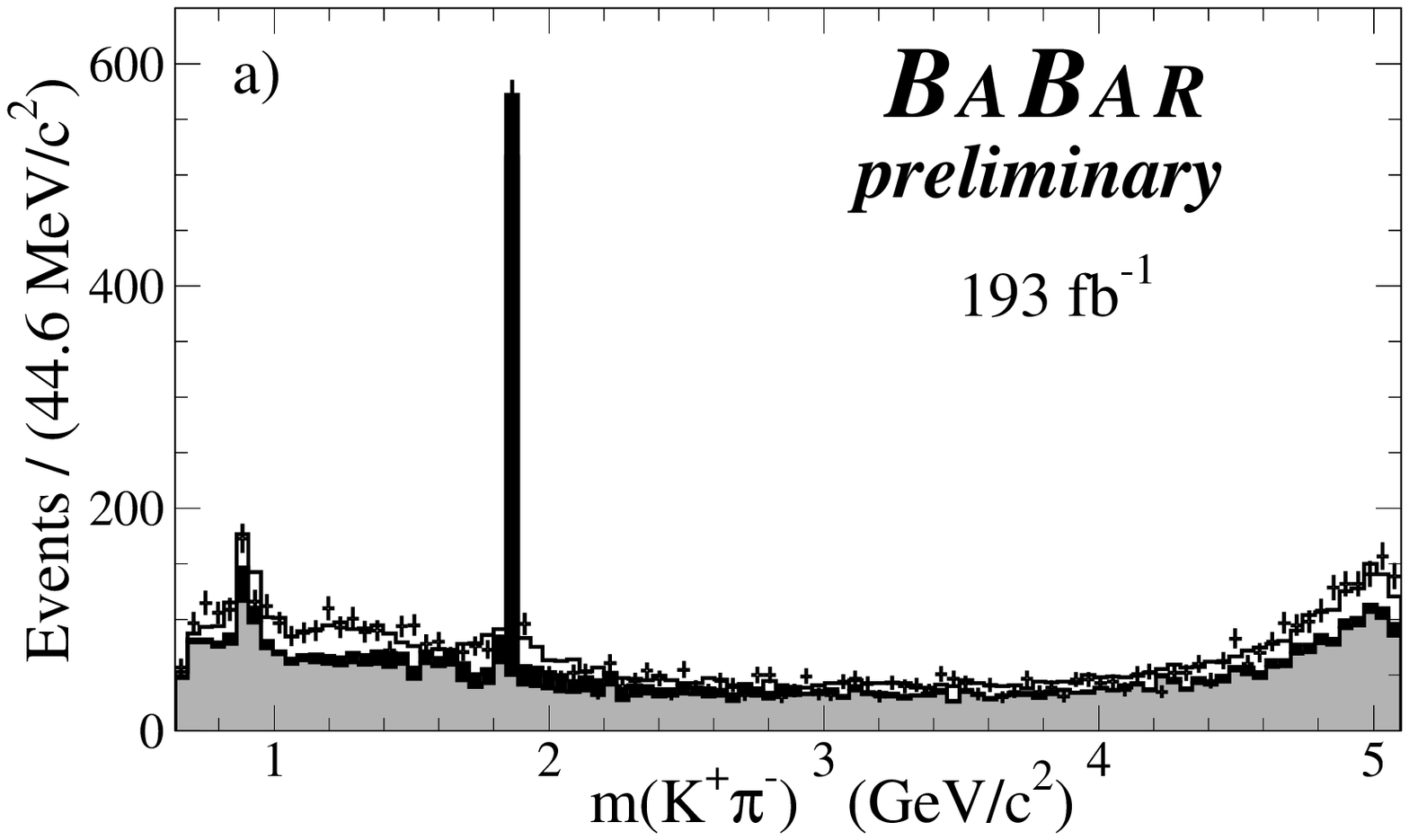}
  \includegraphics[height=.198\textheight]{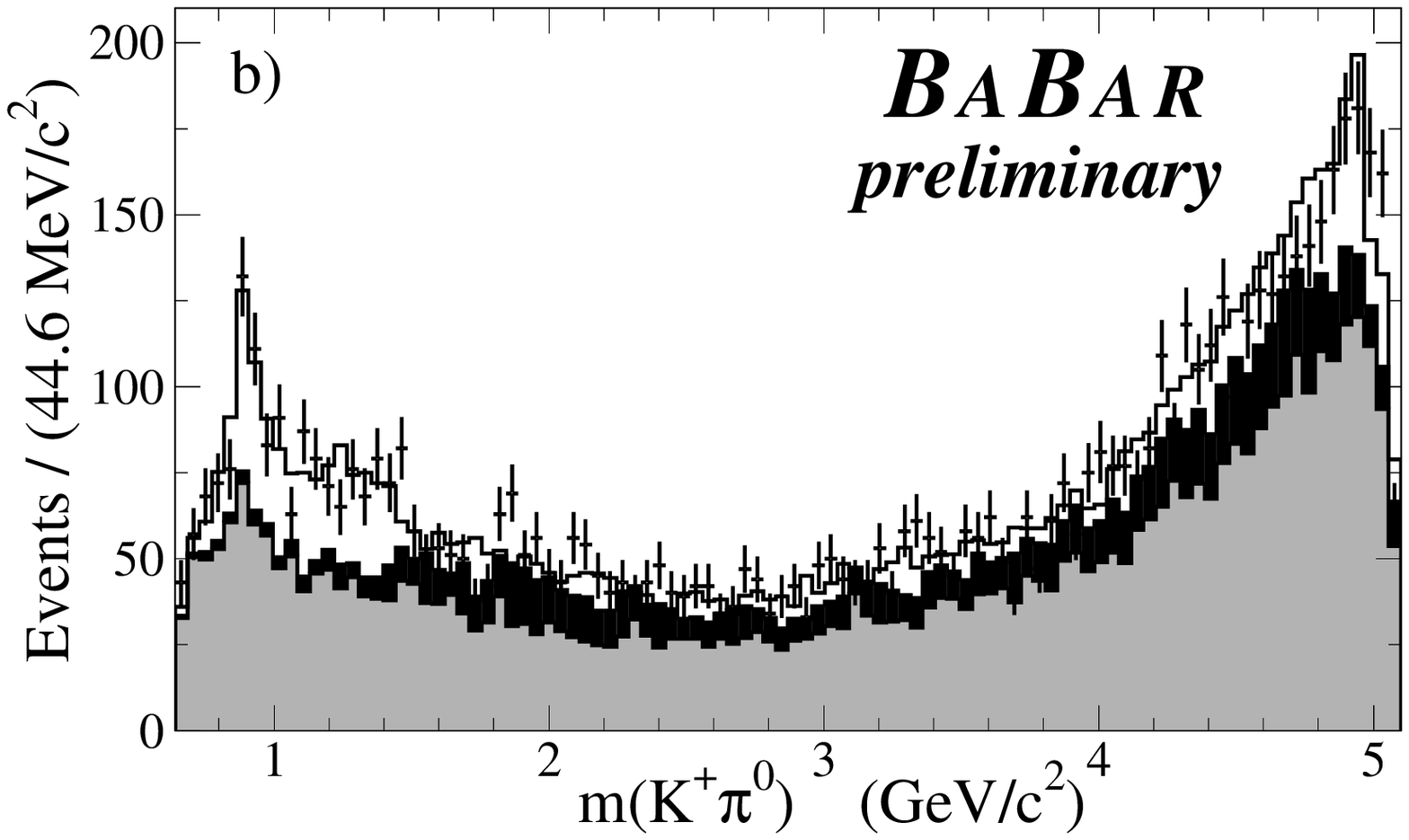}
  \caption{
  \label{dalitz}
  The a) $K^+\pim$ and b) $K^+\pi^0$ invariant mass distributions in
  selected $B^0\to K^+\pim\pi^0$ candidates (points with error bars).
  The gray areas represent the combinatoric background from \epem
  annihilations into hadrons, and the black areas the contributions from other
  $B$ meson decays, including the $B^0 \to \overline{D}^0\pi^0$ 
  intermediate state.
  The histograms represent the result of a fit.
}
\end{figure}

Dalitz analyses of three-body modes can supply a wealth of
spectroscopic information.
We have analyzed several $D$ meson decays with much
higher statistics than earlier experiments,
and are starting to study decays of the much heavier $B$ mesons, both for
their own sake and to maximize CP information.
We have preliminary results for $B^0\!\to K^+\pim\pi^0$, 
$B^+\!\to K^+\pim\pip$ and $B^+\!\to \pip\pim\pip$~\cite{dalhepex}.
The $M(K^+\pim)$ and $M(K^+\pi^0)$ projections
for the former are shown in Fig.~\ref{dalitz}.
The background from $\epem \to q\overline{q}$ events is
large but measurable in off-resonance data.
The background from other $B$ meson decays can be more complicated: 
here we consider $B^0\! \to \overline{D}^0\pi^0$ as background, 
since the $\overline{D}^0$ decays weakly and its mass projections 
are calculable; 
the remaining $B$ decay background is suppressed strongly by the selection
criteria, but must be modelled.

These data yield qualitative information on the internal structure 
of this decay.
Both $K\pi$ systems are dominated by S-wave components,
with $K_0^*(1430)\pi$ amplitudes 3--4 times the $K^*(890)\pi$ amplitudes.
In the $\pim\pi^0$ system there is a substantial $\rho^-$, 
but S-wave resonances should be suppressed by isospin.
In the $B^+\to K^+\pim\pip$ decay we observe S-wave dominance, with
the $K_0^*(1430)^0\pip$ amplitude being $\sim$5 times larger than that
of the $K^*(890)^0\pip$ and the $K^+f_0(980)$ being about twice the
$K^+\rho^0(770)$.
On the other hand, in the $B^+\to \pip\pim\pip$ decay we observe large
contributions from $\rho(770)^0\pip$, $\rho(1450)^0\pip$ and
$f_2(1270)^0\pip$, but no significant S-wave contribution.

\section{$D_s$ Meson Spectroscopy}

The $D_s^+$ mesons are bound states of a `heavy' charmed quark and a
`medium' mass strange antiquark and their spectrum provides an interesting
theoretical testing ground.
Prior to the $B$ factories, four narrow states were known, with masses
of 1968, 2112, 2536 and 2573 \mevcc, and assigned to the expected
narrow states with respective $J^P=0^-, 1^-, 1^+$ and $2^+$, 
although the spins of the latter three states are unmeasured.
The low-mass $0^+$ and $1^+$ states were expected to be wide and to lie
above $DK$ and $D^*K$ thresholds, respectively,
so it was surprising when two narrow states,
$D_{sJ}(2317)$ and $D_{sJ}(2460)$, were discovered in the isospin 
violating decay modes $D_s(1963)\pi^0$ and $D^*_s(2112)\pi^0$.
These discoveries reinvigorated both experiment and theory in this area,
with abundant theoretical speculation that they may
not be simple $c\overline{s}$ states, but perhaps 4-quark states, $DK$
`molecules' or `hybrid' $c\overline{s}g$ states.
It is of great importance to establish their quantum numbers experimentally.

The observation of specific additional states or decay modes can exclude
certain hypotheses, and we have done a number of studies~\cite{DsJdecay}.
In the $D_s(1968)\pi^\pm$ invariant mass spectra we find no evidence
for the neutral or doubly charged states that would be expected as 
isospin partners of any molecular state.
In the inclusive $D_s(1968)\pim\pip$ invariant mass distribution we
observe signals at 2460 and 2536~\mevcc, excluding the $0^+$
hypothesis for those states; 
there is no signal at 2317~\mevcc.

\begin{figure}
  \includegraphics[height=.225\textheight]{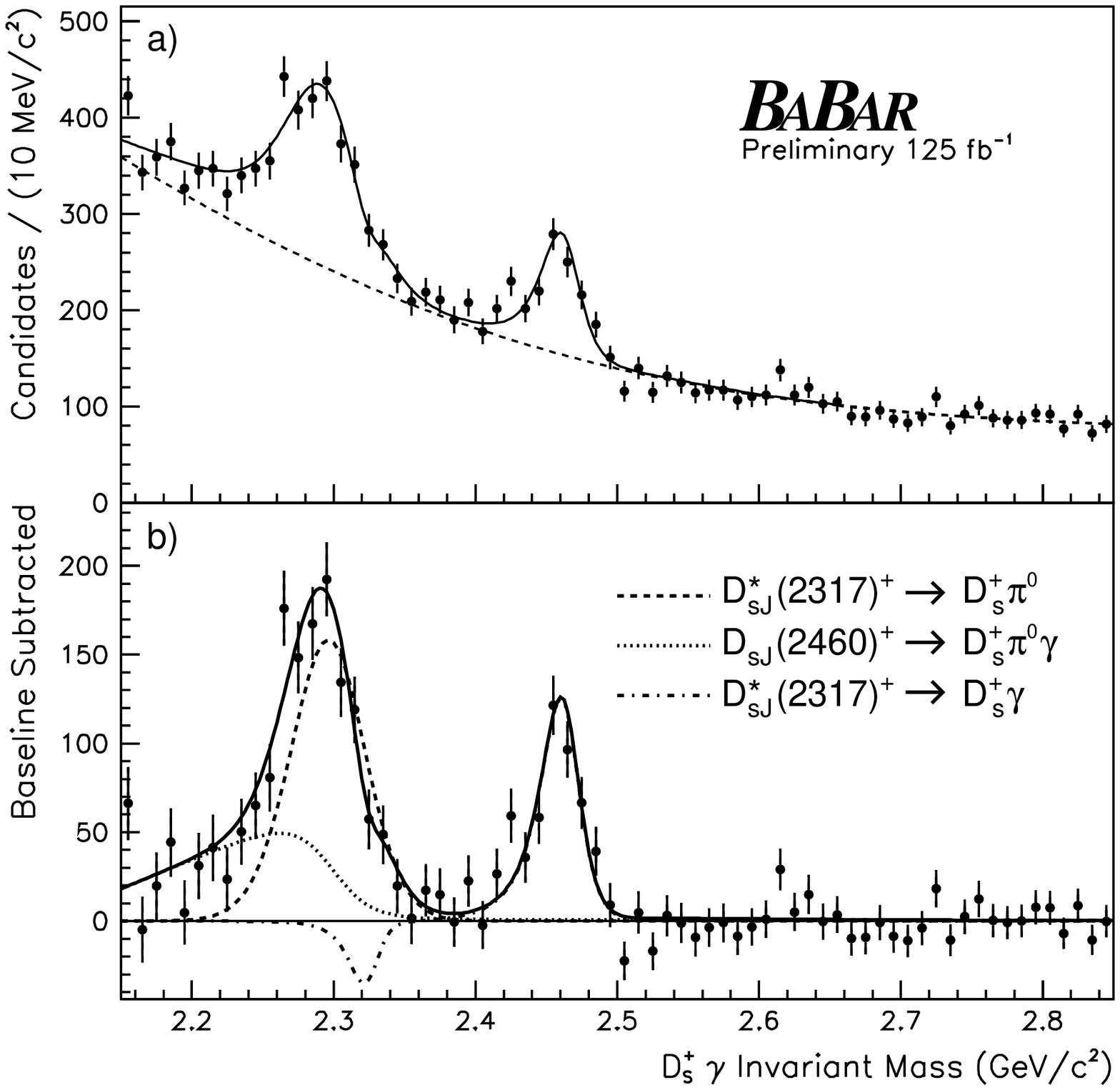}
  \includegraphics[height=.225\textheight]{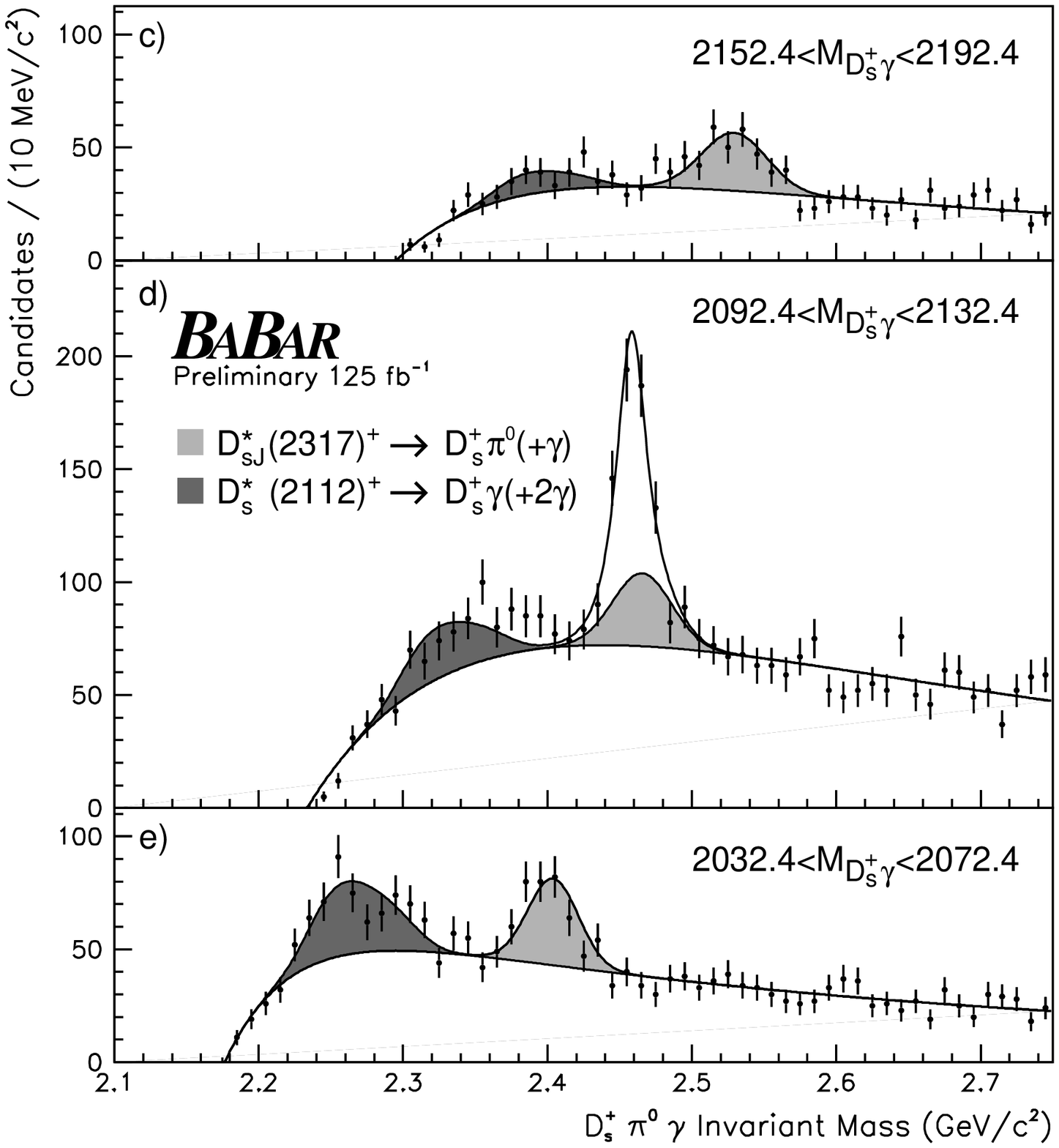}
  \includegraphics[height=.225\textheight]{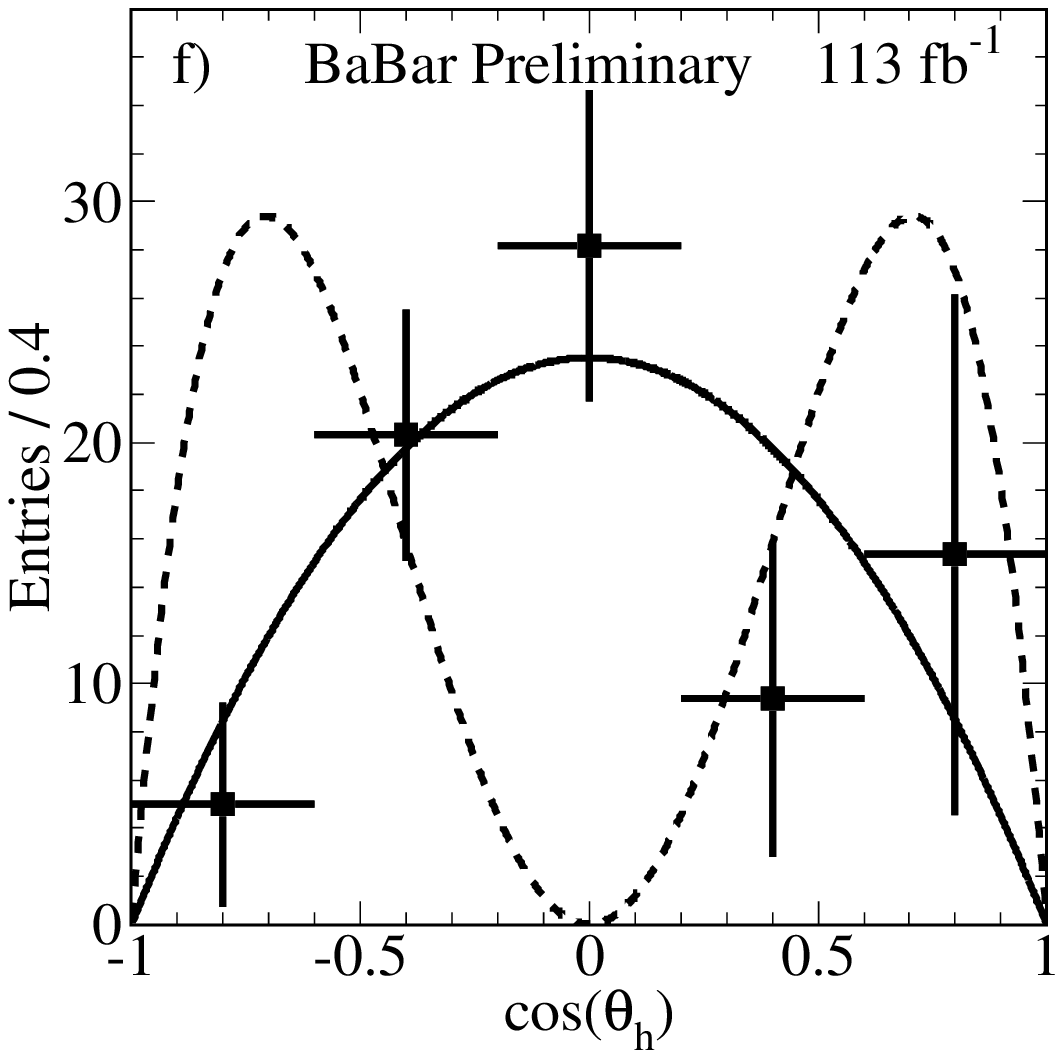}
  \caption{
  \label{DsJplots}
  The inclusive $D_s(1968)\gamma$ invariant mass distribution a) before and
  b) after subtraction of combinatoric background.  The lines
  represent the results of a fit to the data with the components indicated.
  c,d,e) The inclusive $D_s(1968)\pi^0\gamma$ invariant mass distribution 
  in three ranges of the $D_s(1968)\gamma$ submass.
  The lines (gray areas) represent the results of a fit to the data 
  (contributions from the indicated feed-ups).
  f) The distribution of the cosine of the helicity angle for 
  decays of $D_{sJ}(2460)\to D_s \gamma$ in
  $B^0 \to \overline{D}D_{sJ}(2460)$ decays; the
  solid (dashed) line represents the expectation for the spin-1
  (spin-2) hypothesis.
}
\end{figure}

We have searched for radiative decays to $D_s(1963)\gamma$ and 
$D_s(1963)\gamma\pi^0$,
in which there is considerable cross-feed to be taken into account.
In Figs.~\ref{DsJplots}a and~\ref{DsJplots}b we show the inclusive
$D_s^+\gamma$ invariant mass distribution before and after subtraction
of the combinatorial background, respectively, and in
Figs.~\ref{DsJplots}c,d,e we show the inclusive $D_s^+\gamma\pi^0$
invariant mass distribution in three ranges of the $D_s^+\gamma$
submass.
There is a signal for $D_{sJ}(2460)\to D_s^+\gamma$ in
Fig.~\ref{DsJplots}a,b, however the structure at lower mass is due to
a combination of feeddowns from $D_{sJ}(2460)$ decays and 
$D_{sJ}(2317)\to D_s^+\pi^0$ decays where only one of the photons from
the $\pi^0$ is found.
Feed-up from $D_{sJ}(2317)\to D_s^+\pi^0$ and $D_{s}^*(2112)\to D_s^+\gamma$ 
with additional photons attached gives rise to the structures 
in gray in Figs.~\ref{DsJplots}c,d,e that 
move with the the $D_s^+\gamma$ submass.
A signal is seen at 2460~\mevcc only in Fig.~\ref{DsJplots}d, 
indicating that the
$D_{sJ}(2460)\to D_s^+\gamma\pi^0$ decay proceeds almost entirely through the 
$D_{s}^*(2112)\pi^0$ intermediate state.
These data confirm the exclusion of the spin-0 hypothesis for
the $D_{sJ}(2460)$ and suggest the $D_{sJ}(2317)$ has spin 0.

These studies reveal the limitations of working with production in
\eecc events.
We have also studied $D_{sJ}$ production in exclusive $B$
decays~\cite{DsJfromB}, finding evidence for both the 2317 and 2460
states in several $B\to \overline{D}^{(*)}D_{sJ}$ modes.
These signals have lower statistics but also much lower background and
provide complementary information on masses and branching ratios.
In particular we obtain an improved value of
${\cal B}(D_{sJ}(2460)^+ \to D_s\gamma) / 
 {\cal B}(D_{sJ}(2460)^+ \to D_s^*\pi^0) = 0.27\pm0.05$.
Since the $B$ and $D$ are both pseudoscalars, the 
$B\to \overline{D}D_{sJ}$ decays can be used to measure the spin of
the $D_{sJ}$ directly.
The distribution of the cosine of the angle $\theta_h$ between the
$D_s$ flight direction and the $\overline{D}$ flight direction in the 
$D_{sJ}(2460)$ rest frame is shown in Fig~\ref{DsJplots}f, along with
the expectations for the spin-1 and spin-2 hypotheses.
Statistics are low, but the data show a strong preference for spin-1
over spin-2 (or any higher spin).

\section{Pentaquarks}

The claims for the exotic baryonic states \Thp, \Ximm and \Thc 
generated strong interest in \babar\ because of the small widths
and the expectation of a large number of additional narrow states.
In contrast to fixed target experiments,
\epem annihilations produce all members of a given multiplet at about 
the same rate, giving access to more states.
The production rate of such 5-quark states in \epem annihilations is
unknown, 
but ordinary baryons are observed at rates decreasing exponentially
with mass, 
and we observe large samples of baryons with masses above 1540~\mevcc,
so we should be able to provide useful information as to the nature
of these states.

\begin{figure}
  \includegraphics[height=.146\textheight]{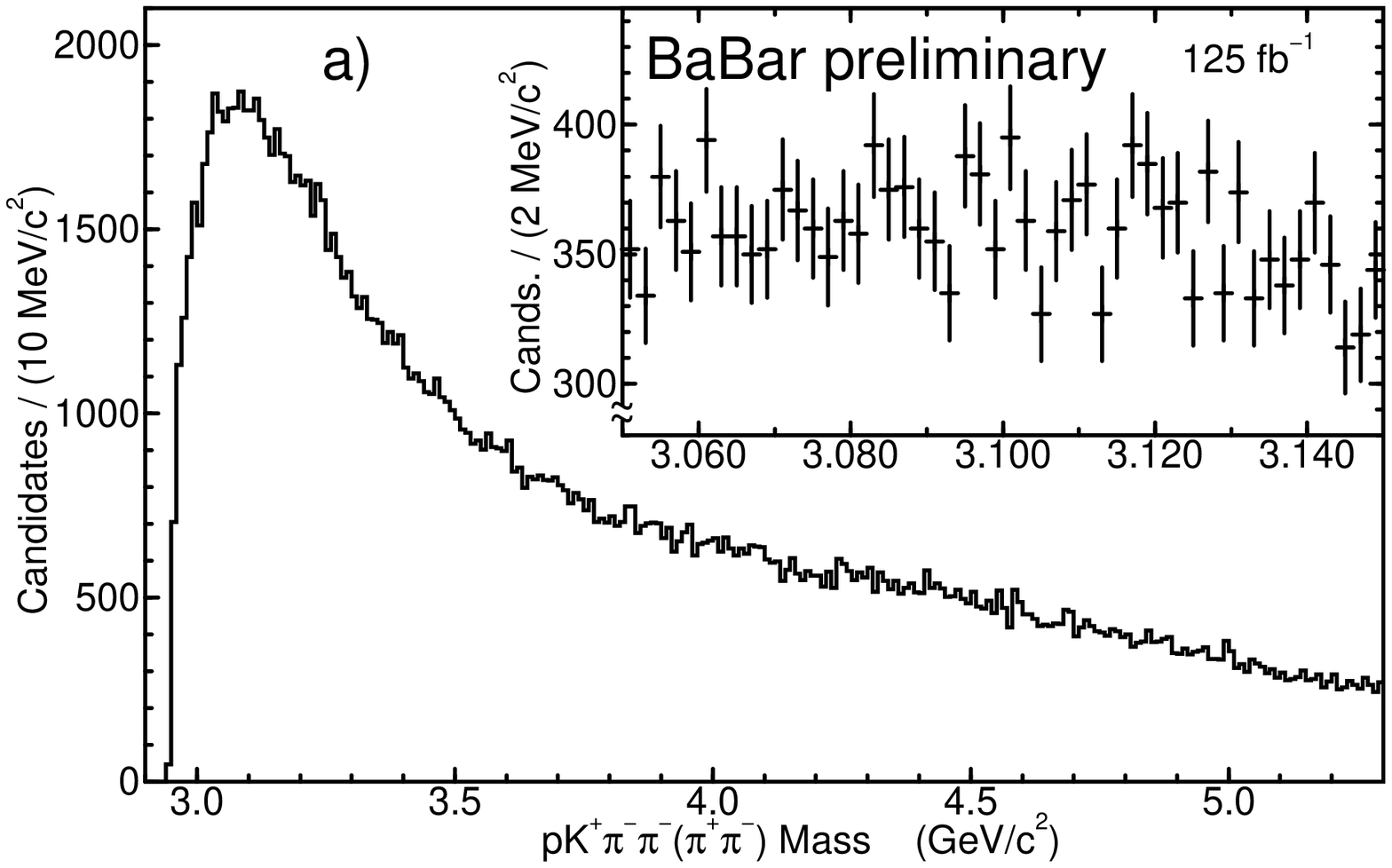}
  \includegraphics[height=.146\textheight]{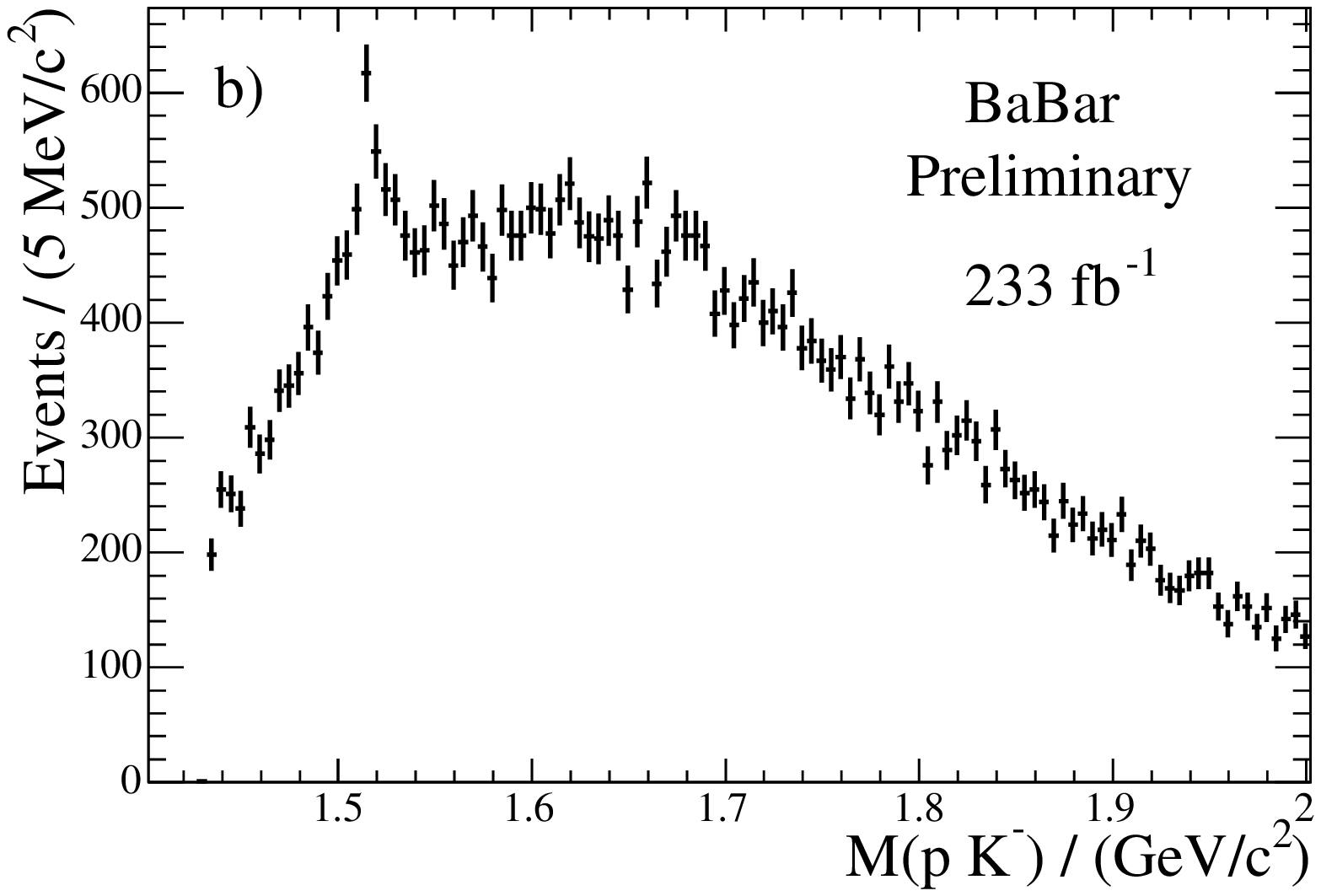}
  \includegraphics[height=.146\textheight]{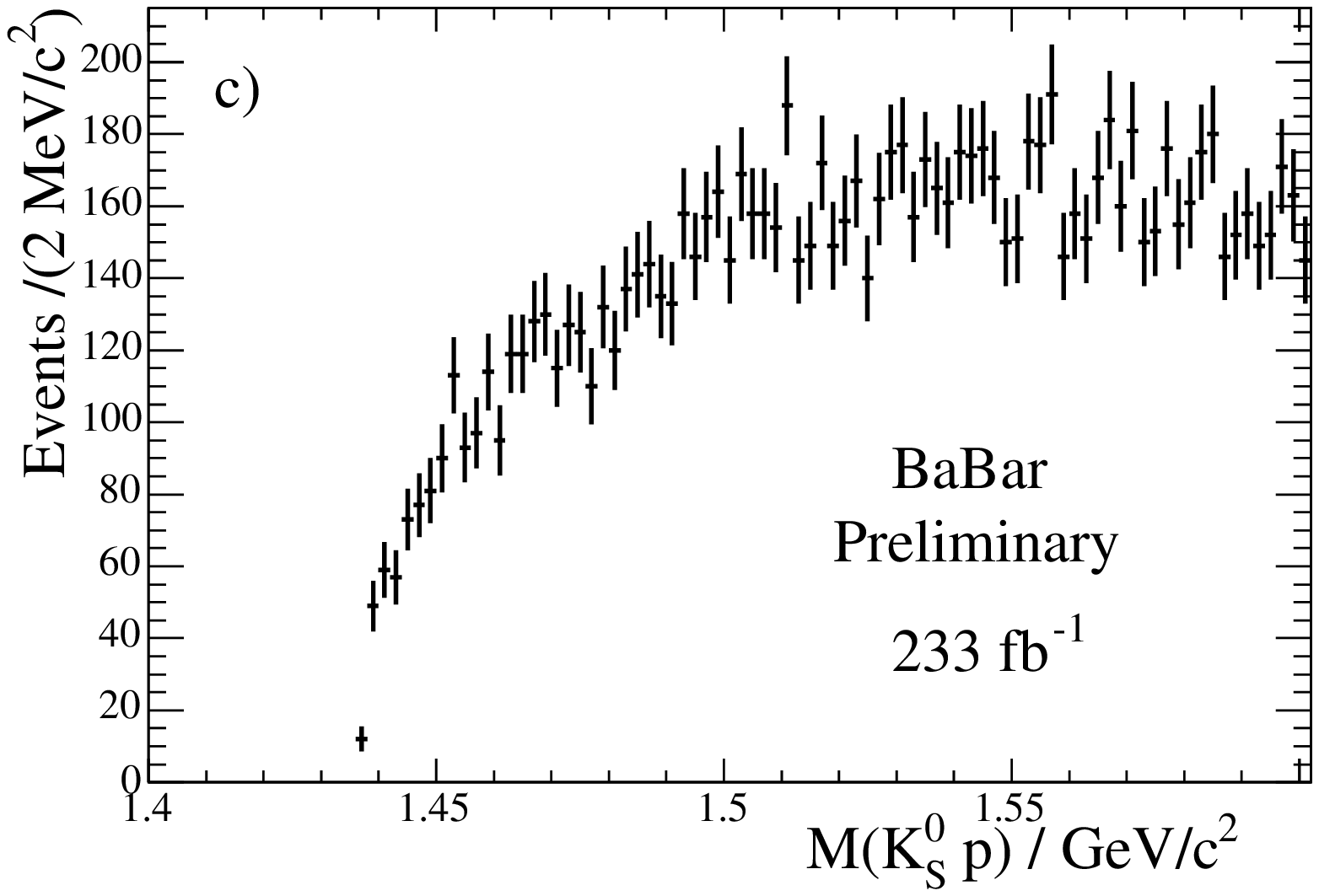}
  \caption{
  \label{pqplots}
  a) The inclusive \pdstm invariant mass distribution over the full
  kinematic range and (inset, with suppressed zero) in the vicinity of
  3100 \mevcc.
  The b) $pK^-$ and c) $p\KS$ invariant mass distributions for $p\KS K^-$ 
  combinations forming a vertex in the beampipe or detector material.
 }
\end{figure}

Extensive searches~\cite{pqincl} in our inclusive \eeqq and \Y4S data 
for the \Thp, \Ximm and other states in the antidecuplet generally 
supposed to contain them yield no signals.
We limit their production rates to factors of 5--15 below the 
corresponding rates for baryons of the same mass, 
strongly suggesting that these states cannot be ordinary baryons.
A preliminary search for the charmed exotic state \Thc claimed by the 
H1 collaboration, using the same \pdstm decay mode, 
is shown in Fig.~\ref{pqplots}a.
Our mass resolution is $\sim$3~\mevcc, and 
there is no evidence for any narrow structure; 
in particular the region near 3100~\mevcc (inset) is smooth.
Few charmed baryon rates have been measured, so we cannot
draw the same conclusion as for the non-charmed states;
however our \dstm sample is two hundred times larger than H1's, so we
can exclude that their signal arises from either of the sources to
which we are sensitive, hard $c$-quark fragmentation or
$B$ hadron decay.

We search for exotic states in several exclusive $B$ hadron decays, 
finding no signal.
For example, in~\cite{pqbdk} we study $B^+\! \to p\overline{p} K^+$
decays, observing a threshold enhancement in $M(p\overline{p})$,
the $\eta_c K^+$, $J/\psi K^+$ and $\psi(2S) K^+$
intermediate states, 
and evidence for the $\chi_c K^+$ and $p\overline{\Lambda}(1520)$ 
intermediate states.
We see no other structure, such as the excess in
$M(pK^+)$ near 1650~\mevcc predicted for an excited
$\Theta$ multiplet, and we set a stringent limit on $\Theta^{*++}$
production in $B$ decays.

These negative results do not exclude these states' existence, 
so we search for hadro- and electro-production of \Thp in our beampipe 
and detector material.
Along with high luminosity come high backgrounds from final state
particles and off-momentum beam particles interacting in this
material.
A study of $p\KS$ vertices gives a detailed inner detector map,
and we can isolate interactions in Be, Si, Ta, Fe and C.
The largest contribution of over 300,000 vertices from $\sim$9~\gev
electrons hitting the Be of the beampipe constitutes an
electroproduction experiment rather similar to several of the
experiments claiming a signal.
An observation here, combined with the non-observation in \epem
annihilations by the same experiment would be decisive.

We see no signal for the \Thp in the $M(p\KS)$ spectrum
for any type of interaction.
We isolate more exclusive reactions by rejecting events in which another 
track identified as a proton, deuteron or triton can be attached to
the $p\KS$ vertex.
In vertices with an associated $\pi^\pm$, we observe
$K^*(890)^+\!\! \to \KS \pip$, $K^*(890)^-\!\! \to \KS \pim$ and 
$\Lambda\!\! \to p \pim$, but no structure in the $M(p\KS)$ distribution.
We consider vertices with an associated $K^-$; if no other
strange particle is produced in such an event, then the \KS must
have been a $K^0$ rather than a $\overline{K}^0$ and the $p\KS$
combination is exotic.
The $M(pK^-)$ distribution in such events is shown in Fig.~\ref{pqplots}b;
a signal for $\Lambda(1520)$ is evident.
The $M(p\KS)$ distribution in the same events, Fig.~\ref{pqplots}c,
shows no signal near 1540~\mevcc.

\section{Charmonium Spectroscopy}

The field of charmonium spectroscopy has also been revitalized with
the observations of the $\eta_c(2S)$ and $h_c$ states below $D\overline{D}$
threshold, 
the discovery of the $X(3872)$ state well above $D\overline{D}$ 
threshold but decaying to $J/\psi\pip\pim$, 
and the discovery of states at 3940~\mevcc in three different 
production processes decaying into non-$D\overline{D}$ modes.
Theoretical models predict numerous wide charmonium states 
above $D\overline{D}$ threshold, 
but the only experimental data, from the total hadronic cross
section in \epem, show only broad structures and the need for at 
least three states.
Again there is speculation regarding molecules, hybrids, etc.
To improve the situation, further experimental searches and 
systematic studies of the spectrum in different production processes
are needed.

\begin{figure}
  \includegraphics[height=.285\textheight]{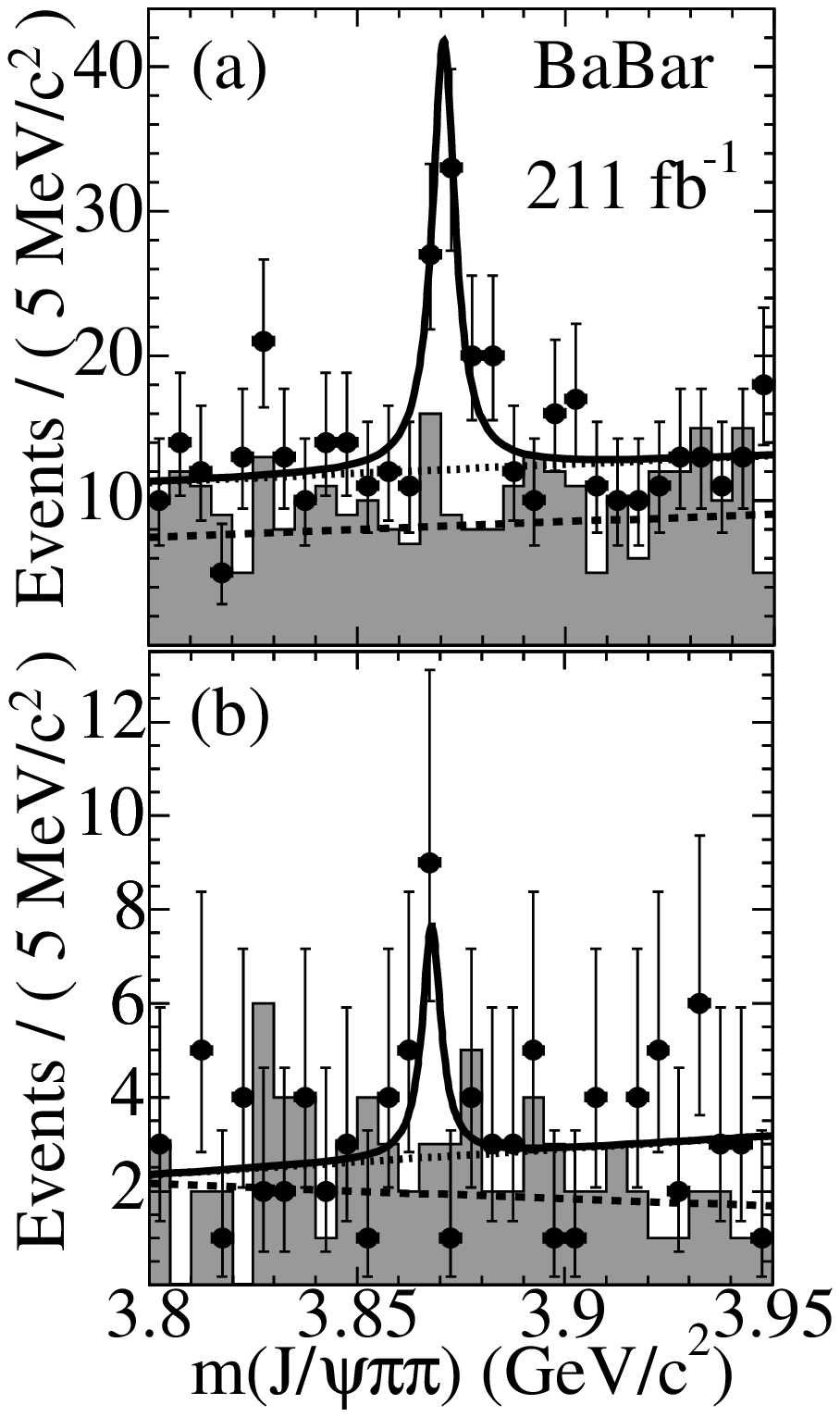}
  \includegraphics[height=.285\textheight]{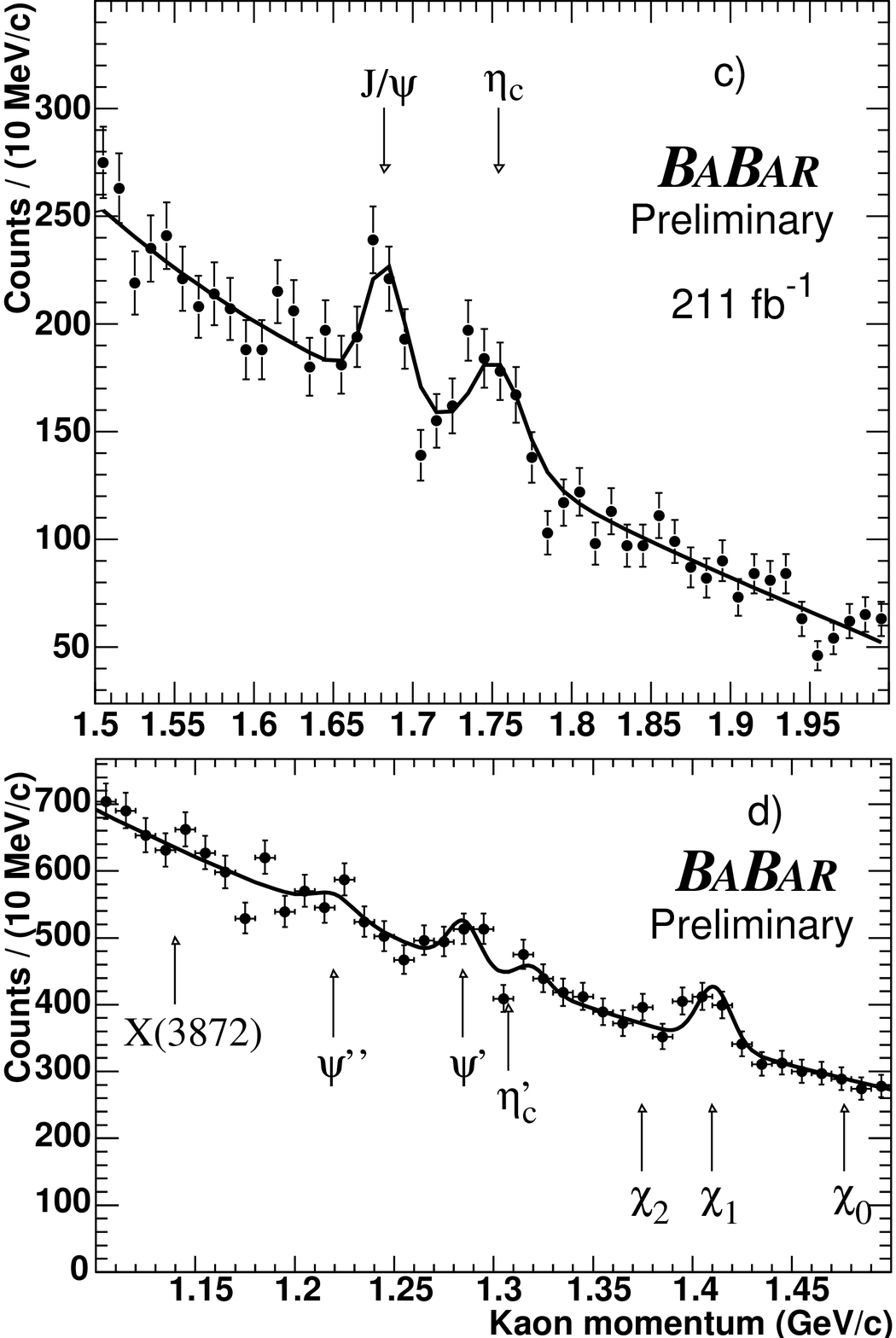}
  \includegraphics[height=.285\textheight]{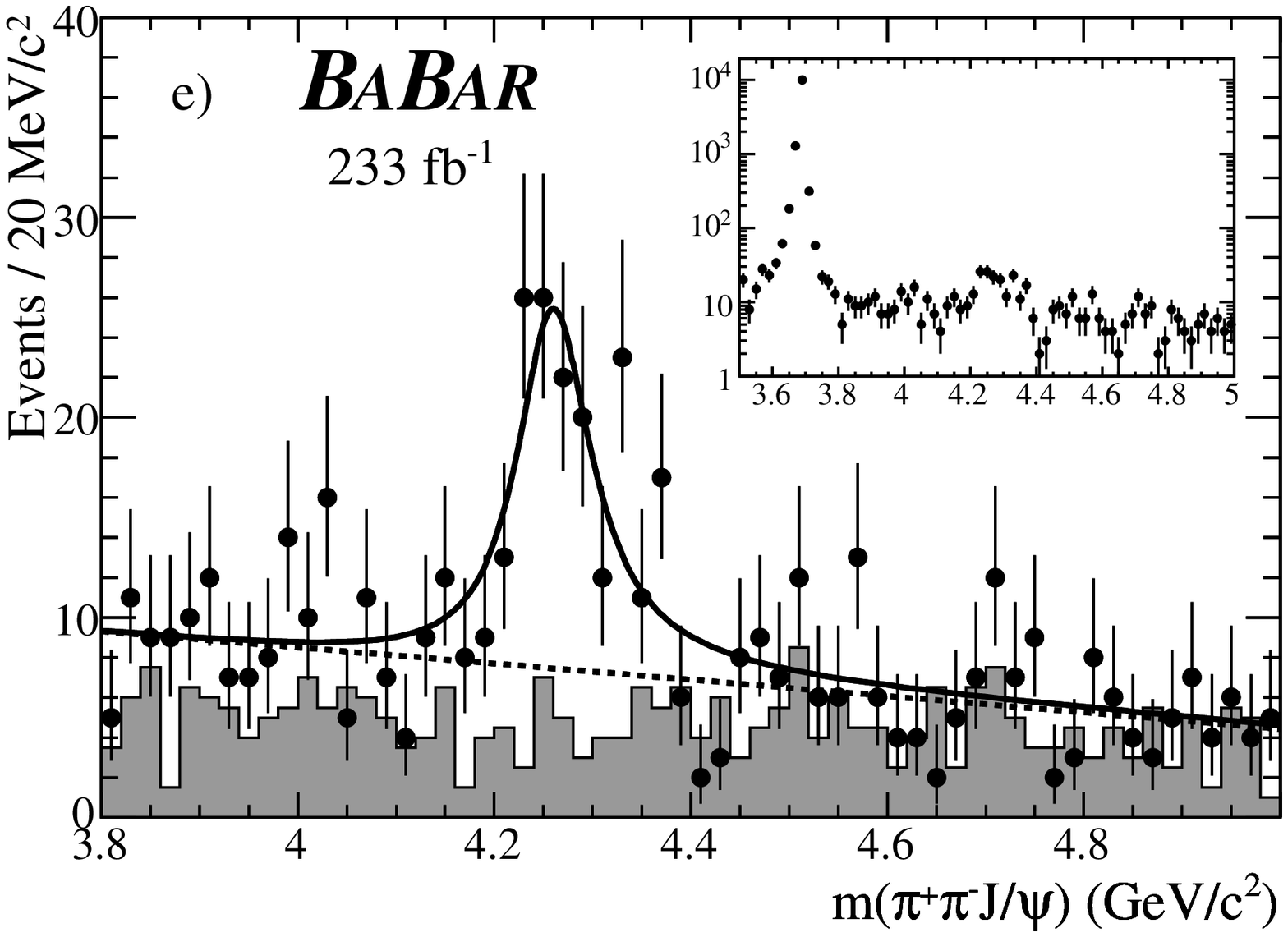}
  \caption{
  \label{chiumplots}
  Invariant mass distributions for the $J/\psi\pip\pim$ system in candidate
  a)~$B^+ \!\to J\psi\pip\pim K^+$ and
  b)~$B^0 \!\to J\psi\pip\pim \KS$ decays, in the vicinity of
  3870~\mevcc.
  The lines represent the results of a fit.
  The $K^+$ momentum spectrum in the rest frame of the system recoiling 
  against a reconstructed $B^-$ candidate, in the ranges corresponding to
  c) $B^+ \to J/\psi K^+$ and $B^+ \to \eta_c K^+$ decays and
  d) $B^+$ decays to a $K^+$ and a higher charmonium resonance.
  e) The $J/\psi\pip\pim$ invariant mass distribution for ISR events
  in the vicinity of 4300~\mevcc and (inset) on a log scale over a
  wider range that includes the $\psi(2S)$.
}
\end{figure}

We set a stringent limit on $X(3872)$ production via ISR~\cite{isr34pi},
strongly disfavoring the $1^{--}$ hypothesis.
We do observe it~\cite{BtoXK} in the decay 
$B^+\! \to X(3872)K^+\! \to J/\psi\pip\pim K^+$, as shown in
Fig.~\ref{chiumplots}a.
The analogous neutral decay
$B^0\! \to X(3872)\KS\! \to J/\psi\pip\pim \pip\pim$ has the
potential to discriminate between a number of models:
if the $X(3872)$ is a conventional $c\overline{c}$ state, the
two decays should have the same rate, although lower experimental
acceptance reduces the expected number of the latter to about ten;
molecular models predict a suppression of the latter decay;
diquark-antidiquark models predict equal rates, but with the
observed $X$ mass shifted by about 8~\mevcc in the latter case.
Our $M(J/\psi\pip\pim)$ distribution~\cite{BtoXK} 
in these decays is shown in Fig.~\ref{chiumplots}b.
Current data are consistent with all three of these hypotheses; the best
fit, indicated by the line, corresponds to a signal
of about seven events with a mass lower by about 3~\mevcc.

In a new study of charmonium from $B$ decays, we search for all 
$B^+\!\to K^+X_{c\overline{c}}$ decays by reconstructing the
accompanying $B^-$ in the event and measuring the $K^+$ momentum
spectrum in the rest frame of the $B^+$.
We thus measure the absolute branching fractions of the $B$ decays, 
independent of the $X_{c\overline{c}}$ decay mode, which can
be used along with existing data to improve our knowledge of a number
of other branching fractions and decay widths.
The kaon momentum spectrum is shown in Figs.~\ref{chiumplots}c,d after 
a selection that enhances quasi-two-body decays and has been 
optimized separately in the two mass regions shown.
There are signals for $X_{c\overline{c}}=J/\psi, \eta_c, \chi_{c1}$, 
and $\psi(2S)$, as well as enhancements corresponding to $\eta_c(2S)$
and $\psi(3770)$.
The observed suppression of $\chi_{c0}$ and $\chi_{c2}$ is predicted
by factorization models.
We see no evidence for $h_c$ production, although the uncertainty is
large due to its proximity to the $\chi_{c1}$.
We see no signal for the $X(3872)$; since we have observed the
corresponding exclusive decay, we set a {\it lower} limit of
${\cal B}(X(3872) \to J/\psi\pip\pim) > 4$\% at the 90\% C.L.
No additional states are observed; the region above the
$X(3872)$ corresponds to $K^+$ momenta below 1.1~\gevc where the
background increases rapidly, and is under study.

We verify~\cite{doublech} the Belle result that the exclusive process 
$\epem\! \to\! X_{c\overline{c}}X^\prime_{c\overline{c}}$, 
where $X_{c\overline{c}}$ and $X^\prime_{c\overline{c}}$ are two
charmonium states, has a much higher rate than expected when
$X_{c\overline{c}}=J/\psi$ or $\psi(2S)$ and 
$X^\prime_{c\overline{c}}=\eta_c$, $\chi_{c0}$ or $\eta_c(2S)$.
In addition we study the exclusive process 
$\epem\to J/\psi\pip\pim$ using ISR~\cite{Yprl};
here we do not require a detected ISR photon since leptonic $J/\psi$
decays plus missing mass requirements produce a clean signal, 
but we verify the expected rate and angular distribution of
associated photons.
The raw number of selected events is shown vs.\ $M(J/\psi\pip\pim)$ 
in Fig.~\ref{chiumplots}e on both linear and (inset) logarithmic 
vertical scales;
in the latter the signal for $\psi(2S)$, a vital control mode, is
prominent; 
in the former there is an excess of events near 4260~\mevcc, 
which we designate the $Y(4260)$.
With the current statistics we are unable to determine if this is a
single resonance or a more complex structure.
If one resonance, its width is $\sim$90~\mev;
such a state is expected to decay strongly
to $D\overline{D}$ and contribute to the total hadronic cross section,
whereas the measured cross section is low in this mass region.

\section{Summary}

The excellent detector performance and high statistics in many
different production processes makes \babar\ an exciting place to do 
spectroscopy.
We have already produced results in a number of areas, including
$\epem$ annihilations at \sqrts from $\pip\pim$ threshold up to 
4.5~\gev using ISR, 
charmed hadrons produced in \eecc events and $B$ hadron decays, 
Dalitz analyses of three-body $B$ and $D$ hadron decays, 
and analyses of interactions in the beampipe and detector material.

With CLEO and Belle, we have revitalized fields of $c\overline{s}$
spectroscopy, 
with the discovery of the $D_{sJ}(2317)$ and $D_{sJ}(2460)$ states,
and charmonium spectroscopy,
with the observation of several new states, most recently our
$Y(4260)$, 
and the development of inclusive methods such as 
$B\!\to KX_{c\overline{c}}$, 
$\epem\! \to X_{c\overline{c}}X^\prime_{c\overline{c}}$, and ISR in
specific final states such as $J/\psi\pip\pim$.
We played a strong role in the pentaquark controversy, and
look forward to a long and fruitful program in these and additional 
exciting areas.

\bibliographystyle{aipproc}

\end{document}